\documentclass[12pt]{article}
 \usepackage[cp1251]{inputenc} 
 \usepackage[english, russian]{babel} 
 \usepackage{amsmath, latexsym, amssymb, bm, array, graphics, eucal,
 amsfonts,}
 \makeatletter
 \renewcommand{\@biblabel}[1]{#1.\hfill}
 \newcommand{\diag}{\rm \diag\, }

 \renewcommand{\Re}{\mathop{\rm Re}}
 \renewcommand{\Im}{\mathop{\rm Im\,}}

\renewcommand{\div}{\mathop{\rm div}}
\newcommand{\mc}[1]{\mathcal{#1}}
 \makeatother
 
\usepackage[dvips]{graphicx}
\usepackage[flushleft,small]{caption2} 

\begin{document}
 \renewcommand{\abstractname}{\, Abstract}
 \thispagestyle{empty}
 \large
 \renewcommand{\refname}{\begin{center}
 REFERENCES\end{center}}
\newcommand{\E}{\mc{E}}
 \makeatother

\begin{center}
\bf   Magnetic susceptibility and Landau diamagnetism of a quantum
collisional Plasmas with arbitrary degree of
degeneration of electronic gas
\end{center}

\begin{center}
  \bf A. V. Latyshev\footnote{$avlatyshev@mail.ru$} and
  A. A. Yushkanov\footnote{$yushkanov@inbox.ru$}
\end{center}\medskip

\begin{abstract}
The kinetic description of magnetic susceptibility and Landau
diamagnetism of quantum collisional plasmas with any degeration
of electronic gas is given. The correct expression of electric
conductivity of quantum collisional plasmas with any degeration
of electronic gas (see  A. V. Latyshev and  A. A. Yushkanov,
{\it Transverse electrical conductivity of a quantum collisional
plasma in the Mermin approach}. - Theor. and Math. Phys., {\bf
175}(1):559--569 (2013)) is used.

electron gas
\end{abstract}

\begin{center}
\bf  Introduction
\end{center}

Magnetisation of electron gas in a weak magnetic fields
compounds of two independent parts (see, for example, \cite {Landau5}):
from the paramagnetic magne\-ti\-sa\-tion connected
with own (spin) magnetic
momentum of elect\-rons ({\it Pauli's para\-mag\-netism}, W. Pauli, 1927)
and from the diamagnetic mag\-ne\-ti\-sation connected with
quantization of
orbital movement of elect\-rons in a magnetic field ({\it
Landau diamagnetism}, L. D. Landau, 1930).

Landau diamagnetism  was considered till now for a gas of the free
elect\-rons. It has been thus shown, that together with original
approach develo\-ped by Landau, expression for diamagnetism of electron
gas can be obtained on the basis of the kinetic approach
\cite {Silin}.

The kinetic method gives opportunity to calculate the trans\-verse
die\-lect\-ric permeability.  On the basis of this quantity its possible
to obtain the value of the diamagnetic response.

However such calculations till now
were carried out only for collisional\-less case. The matter is that
correct expression for the transverse dielec\-t\-ric
permeability of quantum plasma existed till
 now only for gas of the free
electrons. Expression known till now for the transverse dielectric
perme\-abi\-lity in  a collisional case gave incorrect transition to
the classical case \cite{Kliewer}. So this expression  were accordingly
incorrect.

Central result from \cite{Datta} connects the mean orbital
magnetic moment, a thermodynamic property, with the electrical
resistivity, which characte\-ri\-zes transport properties of
material. In this work
was discussed the important problem of  dissipation (collisions)
influence on  Landau dia\-mag\-ne\-tism. The analysis of this problem
is given in the present article with
use of exact expression of transverse conductivity of quantum plasma.

In work \cite{Kumar} is shown that a classical system of charged
particles moving on a finite but unbounded surface (of a sphere)
has a nonzero orbital diamagnetic moment which can be large.
Here is considered a non-degenerate system with the degeneracy
temperarure much smaller than the room temperature, as in the
case of a doped high-mobility semiconductor.

In work \cite{LY}  for the first time the expression   for
the quantum transverse dielectric
permeability of collisional plasma has been derived. The
ob\-tai\-ned in \cite{LY} expression for
trans\-ver\-se dielectric permeability satisfies
to the necessary requirements of com\-pa\-ti\-bility.

In the present work for the first time  with use of correct
expression for the transverse conductivity \cite{LY} the
kinetic description of a magnetic susceptibility of quantum
collisional plasmas with arbitrary degree of degeneration of
electronic gas is given. The formula for
calculation of Landau
diamagnetism for collisional plasmas is deduced.

The kinetic description of a magnetic susceptibility and
Landau dia\-mag\-netism of a quantum collisional degenerate
and Maxwellian plasma was given in our woks \cite{Lat2} and
\cite{Lat3}.

The graphic analysis of a magnetic susceptibility  and
comparison of a magnetic susceptibility Maxwellian and
degenerate plasmas is made.

\begin{center}
  \bf 2. Magnetic susceptibility of quantum collisional plasmas
with arbitrary degree of degeneration of electronic gas
\end{center}

Magnetization vector $\mathbf{M}$ of electron plasma
is connected with current density $\mathbf{j}$ by the following
expression \cite{Landau8}
$$
{\bf j}=c\, {\rm rot}\,\mathbf{M},
$$
where $c$ is the light velocity.

Magnetization vector $\mathbf{M}$ and a magnetic field
strength
$\mathbf{H}=\rm rot \mathbf{A}$ are connected by the expression
$$
{\bf M}=\chi\,{\bf H}=\chi\,{\rm rot}\,{\bf A},
$$
where $\chi$ is the magnetic susceptibility, $\mathbf{A}$ is the
vector potential.

From these two equalities for current density we have
$$
\mathbf{j}={c}\, {\rm rot}\,\mathbf{M}=
c\,\chi {\rm rot}\, \big({\rm rot}\,\mathbf{A}\big)=
c\,\chi\, \big[{\bf \nabla}({\bf \nabla}\cdot{\bf A})-
{\mathbf{\triangle}}\mathbf{A}\big].
$$

Here $\Delta$ is the Laplace operator.

Let the scalar potential is equal to zero.
Vector potential we take ortho\-gonal
to the direction of a wave vector $\mathbf{q}$
($\mathbf{q}\mathbf{A}=0$) in the form of a  harmonic wave
$$
\mathbf{A}(\mathbf{r},t)=\mathbf{A}_0
e^{i(\mathbf{q} \mathbf{r}-\omega t)}.
$$

Such vector field is solenoidal
$$
\div \mathbf{A}=\nabla\mathbf{A}=0.
$$

Hence, for current density we receive equality
$$
{\bf j}=-c\,\chi \Delta \mathbf{A}=c\,\chi\,q^2 \mathbf{A}.
$$

On the other hand, connection of electric field  $\mathbf{E}$ and vector
potential $\mathbf{A}$ is as follows
$$
\mathbf{E}=-\dfrac{1}{c}\dfrac{\partial \mathbf{A}}{\partial t}=
\dfrac{i\omega}{c}\mathbf{A}.
$$
It is leads to the relation
$$
\mathbf{j}=\sigma_{tr}\mathbf{E}=\sigma_{tr}\dfrac{i\omega}{c}
\mathbf{A},
$$
where $\sigma_{tr}$ is the transverse electric conductivity.

For our case  we obtain
following expression for the magnetic sus\-cep\-ti\-bi\-lity
$$
\chi=\dfrac{i\omega}{c^2 q^2}\sigma_{tr}.
\eqno{(1.1)}
$$

Expression of transversal conductivity of collisional
plasmas with arbitrary degree of degeneration of electronic gas
is defined by the general formula \cite{LY}:
$$
\sigma_{tr}({\bf q},\omega,\nu)=\sigma_0\dfrac{i \nu}{\omega}\Big(1+
\dfrac{\omega B({\bf q},\omega+i\nu)+i \nu B({\bf q},0)}
{\omega+i \nu}\Big),
\eqno{(1.2)}
$$
where $\sigma_0$ is the static conductivity,
$\sigma_0={e^2N}/{m\nu}$, $N$ is the concentration (number density)
of plasmas particles, $e$ and $m$ is the electron charge and mass,
$\nu$ is the effective collisional frequency of plasmas particles,
$$
B({\bf q},0)=\dfrac{\hbar^2}{8\pi^3mN}\int
\dfrac{f_{\bf k}-f_{\bf k-q}}
{\E_{\bf k}-\E_{\bf k-q}}{\bf k}_\perp^2d^3k,
$$

$$
B({\bf q},\omega+i\nu)=\dfrac{\hbar^2}{8\pi^3mN}
\int \dfrac{f_{\bf k}-f_{\bf k-q}}
{\E_{\bf k}-\E_{\bf k-q}-\hbar(\omega+i \nu)}{\bf k}_\perp^2d^3k,
$$
$$
f_{\bf k}=\Big[1+\exp \Big(\dfrac{\E_{{\bf k}}-\mu}
{k_BT}\Big)\Big]^{-1},\qquad
\E_{\bf k}=\dfrac{\hbar^2{\bf k}^2}{2m},
$$
$$
\E_T=\dfrac{mv_T^2}{2}=\dfrac{\hbar^2k_T^2}{2m}=k_BT,
$$
$\E_{\bf k}$ is the electrons energy,
$\E_T$ is the heat electrons energy, $k_B$ is the Boltzmann's
constant,
$v_T=1/\sqrt{\beta}$ is the heat electrons velocity,
$\beta=m/2k_BT$,
$\hbar$ is the Planck's constant,

$$
{\bf k}_\perp^2={\bf k}^2-\Big(\dfrac{{\bf kq}}{q}\Big)^2.
$$

According to (1.1) and (1.2) magnetic susceptibility of the quantum
collisional plasmas with arbitrary degree of degeneration of
electronic gas is equal
$$
\chi({\bf q},\omega,\nu)=-\dfrac{e^2N}{mc^2q^2}\Big(1+
\dfrac{\omega B({\bf q},\omega+i\nu)+i \nu B({\bf q},0)}
{\omega+i \nu}\Big).
\eqno{(1.3)}
$$

From the formula (1.3) follows, that at $\omega=0$ frequency
of collisions plasma particles $ \nu $ drops out of the formula (1.3).
Hence, the magnetic susceptibility in a static limit does not depend from
frequencies of colli\-si\-ons of plasma and the following form also has:

$$
\chi({\bf q},0,\nu)=-\dfrac{e^2N}{mc^2q^2}\Big[1+
\dfrac{\hbar^2}{8\pi^3mN}\int \dfrac{f_{\bf k}-f_{\bf k-q}}
{\E_{\bf k}-\E_{\bf k-q}}{\bf k}_\perp^2d^3k\Big].
\eqno{(1.4)}
$$

From expression (1.3) follows also, that a magnetic susceptibility in
collisionless quantum plasma with arbitrary degree of degeneration
of electronic gas is equal
$$
\chi({\bf q},\omega,0)=-\dfrac{e^2N}{mc^2q^2}\Big[1+
\dfrac{\hbar^2}{8\pi^3mN}\int \dfrac{f_{\bf k}-f_{\bf k-q}}
{\E_{\bf k}-\E_{\bf k-q}-\hbar\omega}{\bf k}_\perp^2d^3k\Big].
\eqno{(1.5)}
$$
At $ \omega\to  0$ the formula (1.5) passes in the formula (1.4).

Let's deduce the formula for calculation of a magnetic susceptibility
of quantum collisional plasmas
with arbitrary degree of degeneration of electronic gas.

After obvious linear replacement of variables the formula for integral
$B({\bf q}, \omega+i \nu)$ will be transformed to the form
$$
B({\bf q}, \omega+i \nu)=\dfrac{\hbar^2}{8\pi^3mN}\times $$$$ \times
\int \dfrac{(\E_{\bf k+q}+\E_{\bf k-q}-2\E_{\bf k})
f_{\bf k}{\bf k}_\perp^2d^3k}
{[\E_{\bf k}-\E_{\bf k-q}-\hbar(\omega+i \nu)]
[\E_{\bf k+q}-\E_{\bf k}-\hbar(\omega+i \nu)]}.
\eqno{(1.6)}
$$

Let's enter dimensionless variables
$$
z=\dfrac{\omega+i \nu}{k_Tv_T}=x+iy, \qquad
x=\dfrac{\omega}{k_Tv_T}, \qquad y=\dfrac{\nu}{k_Tv_T},
$$
$$
\alpha=\dfrac{\mu}{k_BT}, \qquad -\infty<\mu<+\infty, \qquad
Q=\dfrac{q}{k_T},
$$
$\alpha$ is the dimensionless (normalized, reduced) chemical
potential, .

Let us pass to integration on the vector ${\bf K} = {\bf k}/k_T$,
where $k_T=p_T/\hbar=mv_T/\hbar $ is the thermal wave number.
Vectors $ {\bf K, k} $ we will direct along an axis $x $, believing $ {\bf
K} =K_x (1,0,0) $ and $ {\bf k} =k (1,0,0) $.

Then
$$
\E_{{\bf k}}=\dfrac{\hbar^2k_T^2}{2m}K^2=\E_TK^2,
$$
$$
\E_{\bf k}-\E_{\bf k-q}-\hbar(\omega+i \nu)=2\E_TQ\Big(K_x-\dfrac{z}{Q}-
\dfrac{Q}{2}\Big),
$$
$$
\E_{\bf k+q}-\E_{\bf k}-\hbar(\omega+i \nu)=2\E_TQ\Big(K_x-\dfrac{z}{Q}+
\dfrac{Q}{2}\Big),
$$
$$
\E_{\bf k+q}+\E_{\bf k-q}-2\E_{\bf k}=2\E_TQ^2,
$$
$$
f_{{\bf k}}=\Big[1+
\exp\Big(\dfrac{\E_{{\bf k}}}{\E_T}-\alpha\Big)\Big]^{-1}=
\Big[1+\exp(K^2-\alpha)\Big]^{-1}.
$$

Let us notice, that for plasma with any degree of degeneration
of electronic gas the numerical density  in an
equilibrium condition is equal
$$
N=\dfrac{f_2(\alpha)}{\pi^2}k_T^3,
$$
where
$$
f_2(\alpha)=\int\limits_{0}^{\infty}x^2f_F(x)dx=\int\limits_{0}^{\infty}
\dfrac{x^2dx}{1+e^{x^2-\alpha}}=\dfrac{1}{2}\int\limits_{0}^{\infty}
\ln(1+e^{\alpha-x^2})dx.
$$

In dimensionless parameters integral $B({\bf q},\omega+i \nu)$
equals
$$
B(Q,z)=\dfrac{1}{8\pi f_2(\alpha)Q}\int \dfrac{(f_{{\bf K}}-f_{{\bf K-Q}})
{\bf K}_\perp^2d^3K}{K_x-z/Q-Q/2},
\eqno{(1.7)}
$$
where
$$
{\bf K}_\perp^2=K_y^2+K_z^2,
$$
besides
$$
B(Q,0)=\dfrac{1}{8\pi f_2(\alpha)Q}\int \dfrac{(f_{{\bf K}}-f_{{\bf K-Q}})
{\bf K}_\perp^2d^3K}{K_x-Q/2}.
$$

Let us notice that the integral (1.6) can be transformed to the
following form
$$
B(Q,z)=\dfrac{1}{8\pi f_2(\alpha)}\int \dfrac{f_{{\bf K}}
{\bf K}_\perp^2d^3K}{(K_x-z/Q)^2-(Q/2)^2}.
\eqno{(1.8)}
$$

Now the magnetic susceptibility (2.3) in dimensionless
paramemers equals
$$
\chi(Q,x,y)=-\dfrac{e^2N}{mc^2k_T^2}\cdot\dfrac{1}{Q^2}\Big(1+
\dfrac{x}{z}B(Q,z)+\dfrac{iy}{z}B(Q,0)\Big).
\eqno{(1.9)}
$$

Integrals (2.7) and (2.8) can be reduced to the one-dimensional.
For this purpose
let us notice, that the internal double integral is equal
$$
\int\limits_{-\infty}^{\infty}\int\limits_{-\infty}^{\infty}
\dfrac{(K_y^2+K_z^2)dK_ydK_z}{1+e^{K_x^2+K_y^2+K_z^2-\alpha}}=
2\pi\int\limits_{0}^{\infty}\dfrac{\rho^3d\rho}{1+e^{K_x^2+\rho^2-\alpha}}=
$$
$$
=2\pi f_3(K_x,\alpha)=2\pi l_1(K_x,\alpha).
$$

Here
$$
f_3(K_x,\alpha)=\int\limits_{0}^{\infty}\dfrac{\rho^3d\rho}
{1+e^{K_x^2+\rho^2-\alpha}}
=\int\limits_{0}^{\infty}\rho^3f_F(K_x,\rho,\alpha)d\rho,
$$
where
$$
f_F(K_x,\rho,\alpha)=\dfrac{1}{1+e^{K_x^2+\rho^2-\alpha}},
$$
and
$$
l_1(K_x,\alpha)=\int\limits_{0}^{\infty}\rho\ln(1+e^{\alpha-K_x^2-\rho^2}).
$$

Thus, integrals (2.7) and (2.8) can be calculated
to  following formulas
$$
B(Q,z)=\dfrac{1}{4f_2(\alpha)}\int\limits_{-\infty}^{\infty}
\dfrac{f_3(\tau,\alpha)d\tau}{(\tau-z/Q)^2-(Q/2)^2},
$$
$$
B(Q,z)=\dfrac{1}{4f_2(\alpha)Q}\int\limits_{-\infty}^{\infty}
\dfrac{[f_3(\tau,\alpha)-f_3(\tau-Q,\alpha)]d\tau}{\tau-z/Q-Q/2}.
$$

\begin{center}
\bf 3. Landau diamagnetism of quantum collisional plasmas with
arbitrary degree of degeneration of electronic gas
\end{center}

Landau diamagnetism  in collisionless  plasma is usually
 defined as a magnetic susceptibility in a static limit
for a homogeneous external magnetic field.
Thus the diamagnetism value can be found by means of (1.1)
through two  non-commutative limits
$$
\chi_L=\lim\limits_{q\to 0}\Big[\lim\limits_{\omega\to 0}^{}
\chi(q,\omega,\nu=0)\Big].
\eqno{(2.1)}
$$

Here $\nu$ is the effective frequency of electrons with plasma
particles.

At $y=0$ from formula (1.9) we have
$$
\chi(Q,x,0)=-\dfrac{e^2N}{mc^2k_T^2}\cdot\dfrac{1}{Q^2}\times
$$
$$ \times\Bigg(1+
\dfrac{1}{8\pi f_2(\alpha)}\int \dfrac{(f_{{\bf K}}-f_{{\bf K-Q}})
{\bf K}_\perp^2d^3K}{K_x-x/Q-Q/2}\Bigg).
\eqno{(2.2)}
$$

On the basis of (2.1) and (2.2) for Landau diamagnetism we receive
$$
\chi_L=-\dfrac{e^2N}{mc^2k_T^2}\lim\limits_{Q\to 0}\dfrac{1}{Q^2}\times
$$
$$
\times\Bigg[1-\dfrac{1}{8\pi f_2(\alpha)Q}\int
\dfrac{(f_{{\bf K-Q}}-f_{{\bf K}})
{\bf K}_\perp^2d^3K}{K_x-Q/2} \Bigg].
\eqno{(2.3)}
$$

The function
$$
\varphi(Q)=\dfrac{f_{{\bf K-Q}}-f_{{\bf K}}}{K_x-Q/2}
$$
we will expand on degrees $Q $ to the third order inclusive
$$
\varphi(Q)=\dfrac{2K_xe^{K^2-\alpha}}{(1+e^{K^2-\alpha})^2}Q+
\Big[-\dfrac{2K_xe^{K^2-\alpha}}{(1+e^{K^2-\alpha})^2}+
\dfrac{4K_xe^{2(K^2-\alpha)}}{(1+e^{K^2-\alpha})^3}\Big]Q^2+
$$
$$
+\Big[\dfrac{(8K_x^2+6)e^{K^2-\alpha}}{(1+e^{K^2-\alpha})^2}-
\dfrac{(48K_x^2+12)e^{2(K^2-\alpha)}}{(1+e^{K^2-\alpha})^3}+
\dfrac{48K_x^2e^{3(K^2-\alpha)}}{(1+e^{K^2-\alpha})^4}\Big]
\dfrac{Q^3}{6}+\cdots.
$$

Let us substitute this decomposition in (2.3). We will notice,
that a free member in the received equality is equal to zero
$$
1-\dfrac{1}{8\pi f_2(\alpha)}
\int\dfrac{2e^{K^2-\alpha}{\bf K}_\perp^2d^3K}
{(1+e^{K^2-\alpha})}=1-\dfrac{2}{3f_2(\alpha)}\int\limits_{0}^{\infty}
\dfrac{e^{K^2-\alpha}K^4dK}{(1+e^{K^2-\alpha})^2}=0.
$$

Further, the integral at the first degree $Q$ is equal to zero
as integral from odd function on the real axis.

Therefore, Landau diamagnetism equals
$$
\chi_L=\dfrac{e^2N}{6mc^2k_T^2}\cdot \dfrac{1}{8\pi f_2(\alpha)}\int
\Bigg[\dfrac{(8K_x^2+6)e^{K^2-\alpha}}{(1+e^{K^2-\alpha})^2}-$$$$-
\dfrac{(48K_x^2+12)e^{2(K^2-\alpha)}}{(1+e^{K^2-\alpha})^3}+
\dfrac{48K_x^2e^{3(K^2-\alpha)}}{(1+e^{K^2-\alpha})^4}\Bigg]
{\bf K}_\perp^2d^3K,
$$
whence
$$
\chi_L=-\dfrac{e^2N}{12mc^2k_T^2}\cdot\dfrac{f_0(\alpha)}{f_2(\alpha)}.
\eqno{(2.4)}
$$

Here
$$
f_0(\alpha)=\int\limits_{0}^{\infty}f_F(x,\alpha)dx=
\int\limits_{0}^{\infty}\dfrac{dx}{1+e^{x^2-\alpha}}.
$$

In monograph \cite{Anselm} formula of Landau diamagnetism
for quantum collisional plasma
is given in the form, extremely inconvenient to
com\-pa\-ri\-son with our result
(2.4).

Let us pass in the formula (2.4) to the limit, when $\alpha\to-\infty $.
In this case plasma with any degree of degeneration
of electronic gas passes to Maxwellian plasma. In this limit from (2.4)
we receive
$$
\chi_L =-\dfrac{e^2N}{6mc^2k_T^2},
$$
that in accuracy coincides with result from \cite {Lat2}.

Having divided (1.9) on (2.4), we receive expression for the relative
magnetic susceptibility of quantum collisional plasmas with
any degree of degeneration of electronic gas
$$
\dfrac{\chi(Q,x,y)}{\chi_L}=\dfrac{12f_2(\alpha)}{f_0(\alpha)Q^2}
\Big(1+\dfrac{x}{z}B(Q,z)+\dfrac{iy}{z}B(Q,0)\Big).
\eqno{(2.5)}
$$

For graphic research of a magnetic susceptibility we will be
to use the formula (2.5).

On Fig. 1 comparison of magnetic susceptibilities
of degenerate plasmas (curve 1), Maxwellian plasmas (curve 2)
and plasmas with any degree of degeneration in the case when
dimensionless chemical potential $ \alpha=0$
(curve 3) is presented.

From the Fig. 1 follows, that into quantum collisionless plasma
($ \nu=0$) in the static limit ($ \omega=0$) the magnetic
susceptibility is function monotonously decreasing to zero
as function of wave number.

On Fig. 2 dependence of a magnetic susceptibility
collisionless plasmas with any degree of degeneration of
electron gas
from wave number at various values of chemical potential is presented.
Curves 1,2 and 3 correspond to values of the dimensionless chemical
potential $ \alpha=0,-2$ and $2$. With growth of chemical potential
values of the magnetic susceptibility grow also.
From the Fig. 2 we see, that the magnetic susceptibility
is monotonously decreasing
function of wave number at all values of the dimensionless
chemical potential.

On the Figs. 3 and 4 dependences of the magnetic susceptibility
as functi\-ons of dimensionless wave number (Fig. 3), and functi\-ons
of dimensionless frequency of oscillations of an electromagnetic
field (Fig. 4) are presented.

On the Figs. 3 and 4 dependences of the magnetic susceptibility
as functi\-ons on dimensionless wave number (Fig. 3), and functions
on dimension\-less frequency of oscillations of an electromagnetic
field (Fig. 4) are presented.

From the Fig. 3 we see, that the magnetic susceptibility is
monotono\-us\-ly decreasing function on wave number at all values
of dimensionless chemical potential and frequency of oscillations
of electromagnetic field.

\begin{figure}[h]
\begin{flushleft}
\includegraphics[width=15.0cm, height=10cm]{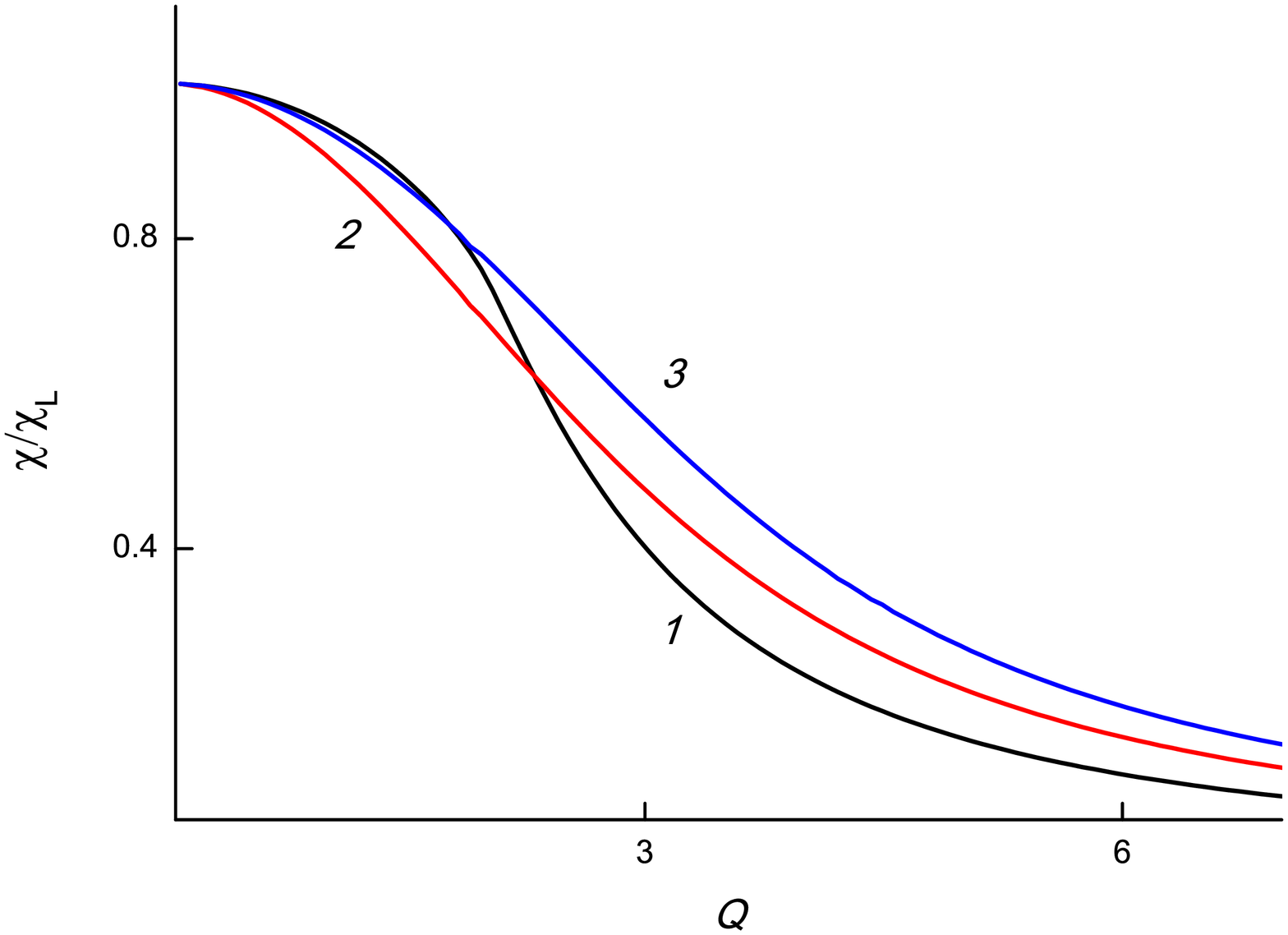}
{Fig. 1. Magnetic susceptibility in static limit ($\omega=0$)
for collisionless plasmas ($\nu=0$). Curve 1 corresponds to
degenerate plasmas, curve 2 corresponds to Maxwellian plasmas,
curve 3 corresponds to plasmas with arbitrary degree of degeneration
of electronic gas at value of dimensionless chemical potential $\alpha=0$.
}\label{rateIII}
\end{flushleft}
\begin{flushleft}
\includegraphics[width=15.0cm, height=10cm]{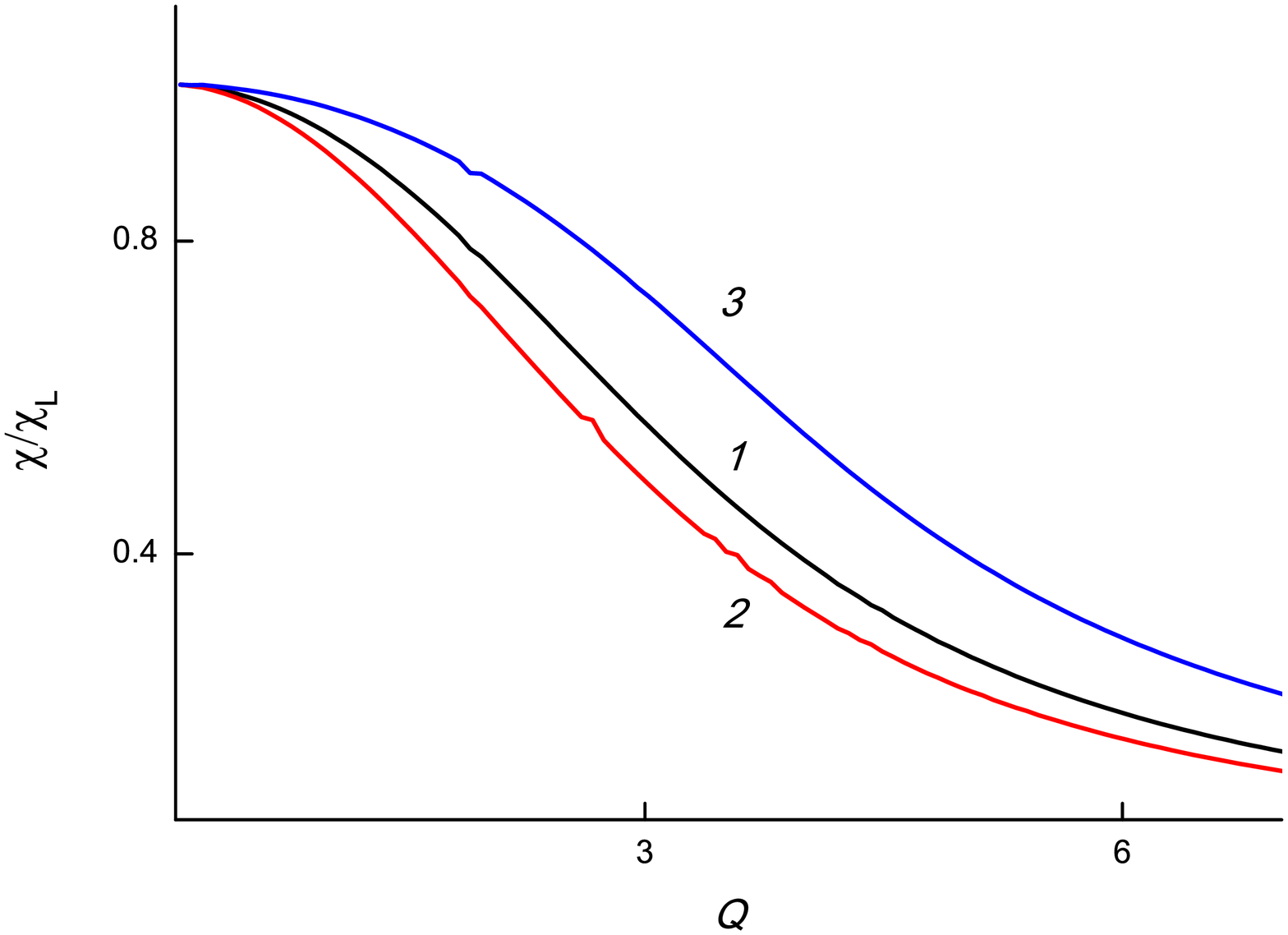}
{Fig. 2. Magnetic susceptibility of collisionless plasmas
($\nu=0$) with arbitrary degree of degeneration
of electronic gas; curves $1$,$2$ and $3$ corresponds to
parameter values $\alpha=0, -2$ and $2$.
}\label{rateIII}
\end{flushleft}
\end{figure}
\clearpage

\begin{figure}[h]
\begin{flushleft}
\includegraphics[width=15.0cm, height=10cm]{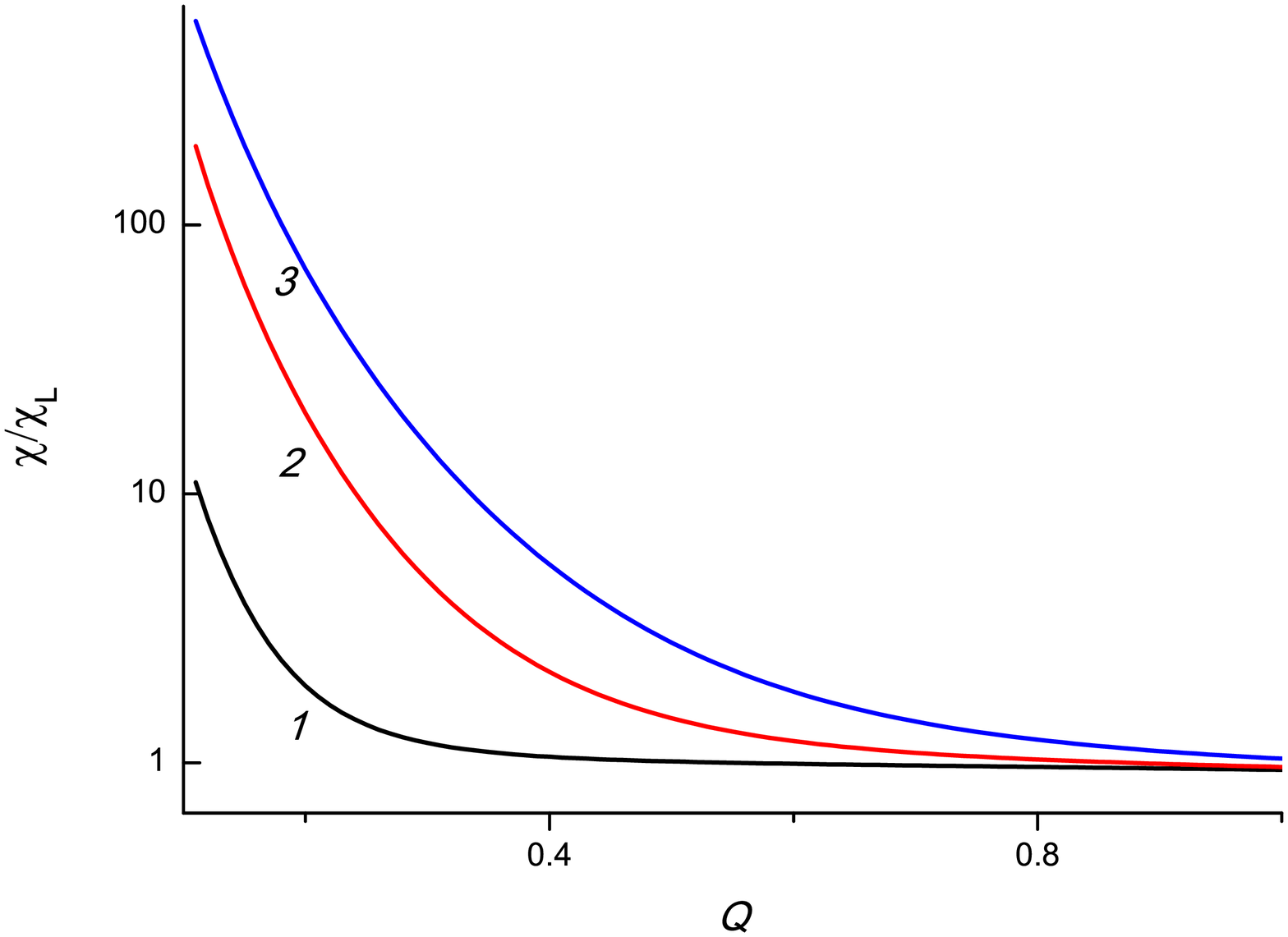}
{Fig. 3. Magnetic susceptibility of collisionless plasmas
($\nu=0$) with arbitrary degree of degeneration
of electronic gas in the case $\alpha=0$;
curves $1$,$2$ and $3$ corresponds to
parameter values $x=0.001, 0.05$ and $0.1$.
}\label{rateIII}
\end{flushleft}
\begin{flushleft}
\includegraphics[width=15.0cm, height=10cm]{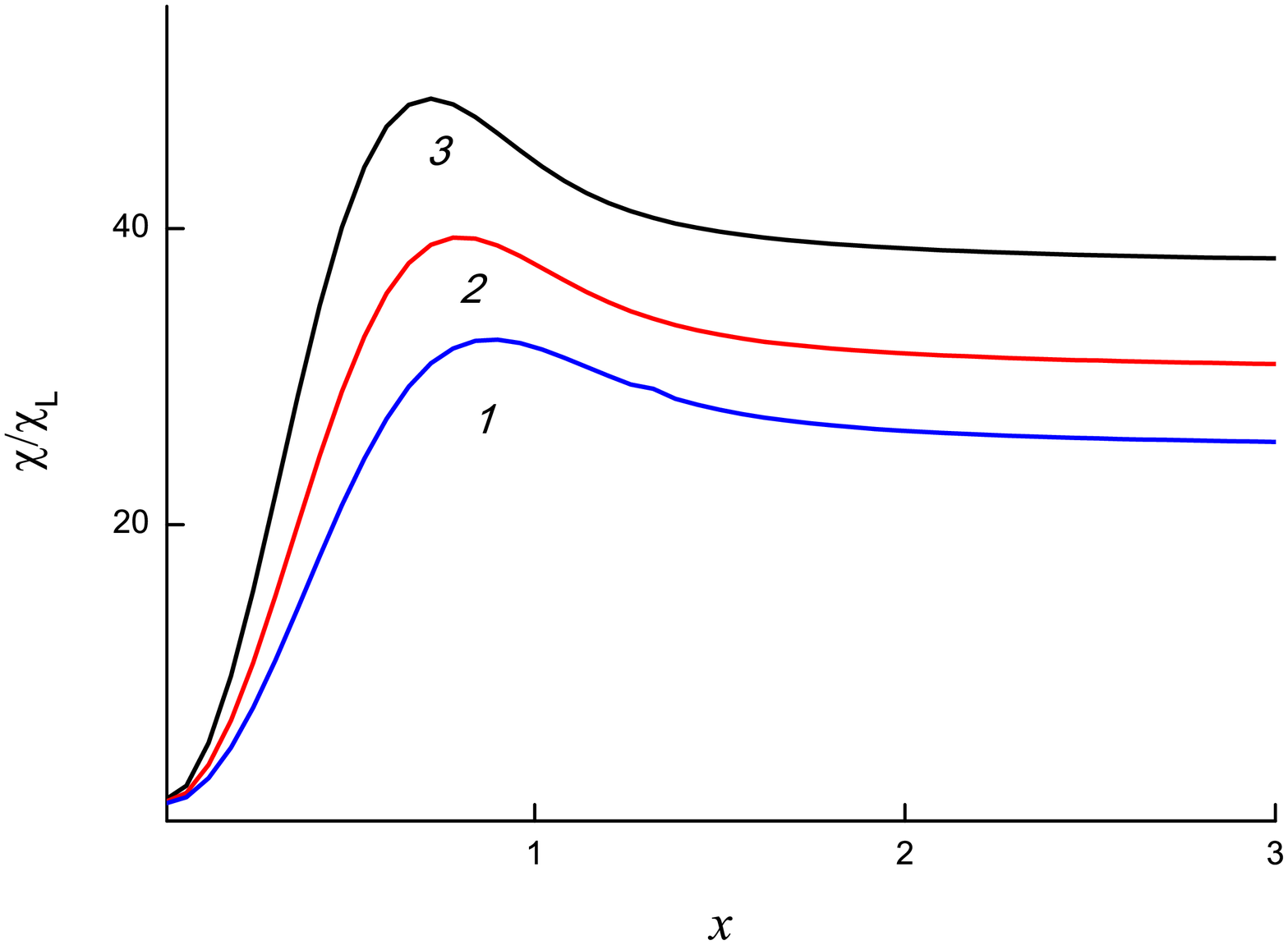}
{Fig. 4. Magnetic susceptibility of collisionless plasmas
($\nu=0$) with arbitrary degree of degeneration
of electronic gas in the case $\alpha=0$;
curves $1$,$2$ and $3$ corresponds to
parameter values   $Q=0.45, 0.5$ and $0.55$.
}\label{rateIII}
\end{flushleft}
\end{figure}
\clearpage

\begin{figure}[h]
\begin{flushleft}
\includegraphics[width=15.0cm, height=10cm]{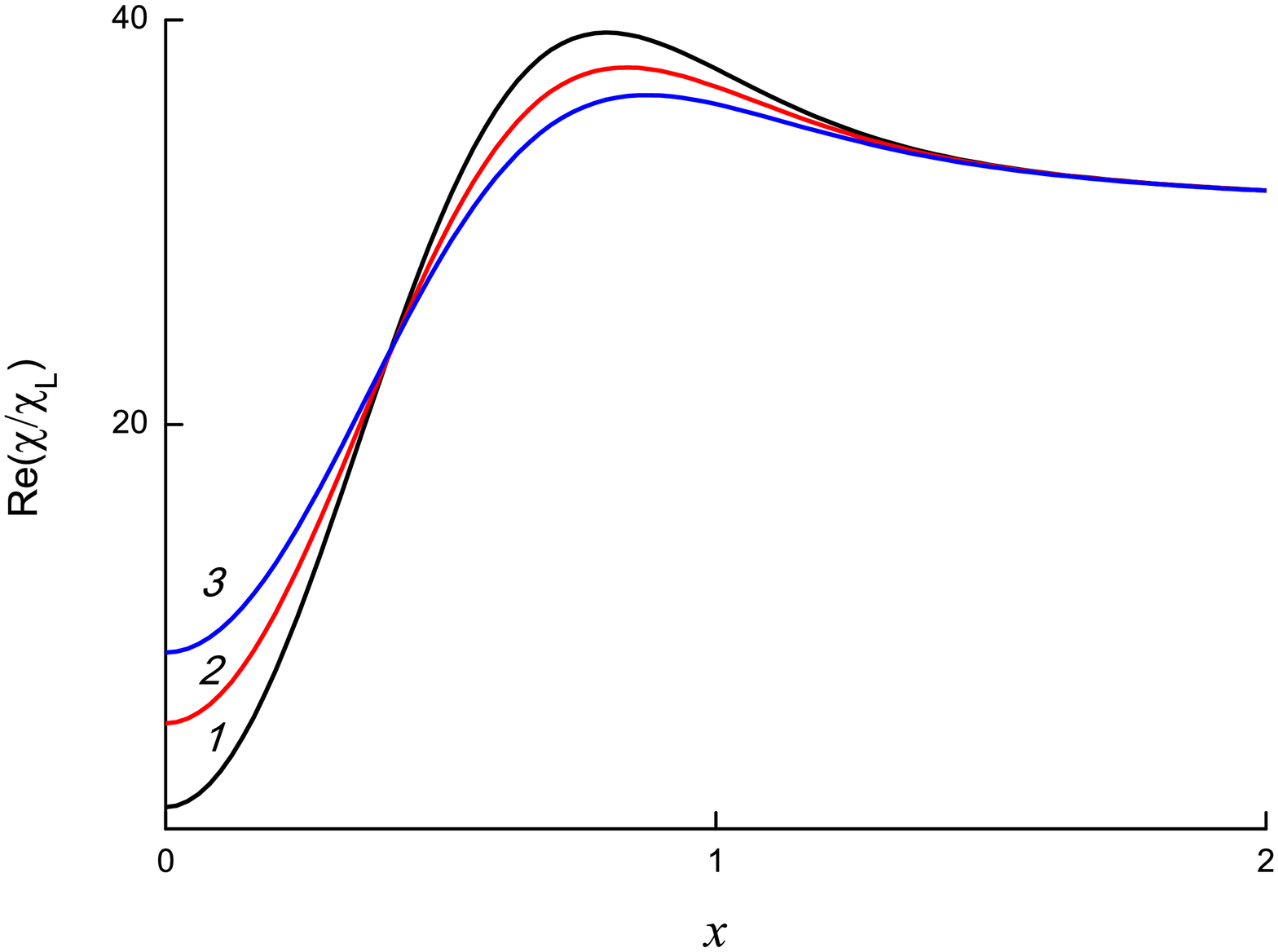}
{Fig. 5. Real part of magnetic susceptibility of quantum plasmas
with arbitrary degree of degeneration
of electronic gas in the case $Q=0.5$ and $\alpha=0$;
curves $1$,$2$ and $3$ corresponds to
parameter values  $y=0.001, 0.05$ и $0.1$.
}\label{rateIII}
\end{flushleft}
\begin{flushleft}
\includegraphics[width=15.0cm, height=10cm]{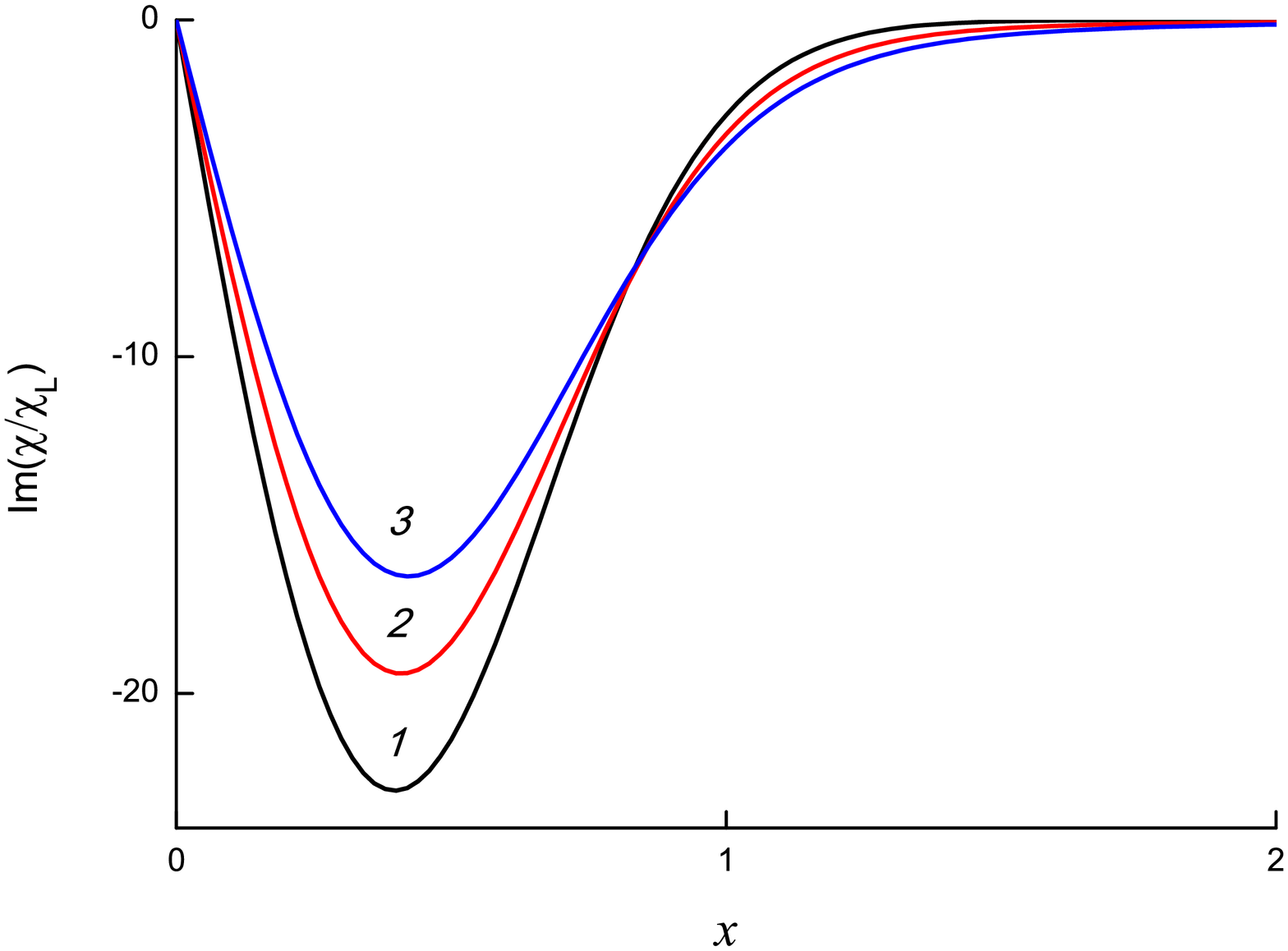}
{Fig. 6. Imaginary part of magnetic susceptibility of quantum plasmas
with arbitrary degree of degeneration
of electronic gas in the case $Q=0.5$ and $\alpha=0$;
curves $1$,$2$ and $3$ corresponds to
parameter values  $y=0.001, 0.05$ и $0.1$.
}\label{rateIII}
\end{flushleft}
\end{figure}
\clearpage

With growth of frequency of oscillations
of electromagnetic field valu\-es of the magnetic
susceptibilities  grow also.

From the Fig. 4 we see, that the magnetic susceptibility
as function of dimensionless frequency of oscillations of
the electromagnetic fields has a maximum and at great values $x $
leaves on the asymptotics
$$
\dfrac {\chi_{\rm as}}{\chi_L} =
\dfrac {12f_2(\alpha)}{f_0 (\alpha) Q^2}.
$$

With growth of wave number values of the magnetic
susceptibi\-li\-ti\-es grow also.

On the Fig. 5 and 6 dependences of real (Fig. 5)  and
imaginary (Fig. 6) parts of magnetic susceptibility
of collisional plasmas as functions on dimensionless
frequencies of oscillations of the electromagnetic
fields in the case $Q=0.5$ are presented.

From the Fig. 5 we see, that the real part of magnetic
susceptibili\-ti\-es  has the maximum and at big $x $ leaves from above
on the asymptotics
$$
\dfrac{\Re\chi_{\rm as}}{\chi_L} =
\dfrac{12f_2 (\alpha} {f_0 (\alpha) Q^2}.
$$

From the Fig. 6 we see, that an imaginary part of magnetic
suscepti\-bi\-li\-ty has a minimum and then leaves from below on the
asymptotics
$$
\dfrac{\Im\chi_{as}}{\chi_L} =0.
$$

Let us notice, that the minimum moves to the right with growth
$Q $. With growth
frequencies of oscillations of an electromagnetic
field the imaginary part tends to zero.

\begin{center}
  \bf 4. Conclusion
\end{center}

In the present work the kinetic description
magnetic susceptibility of quantum collisional plasmas
with any degree of degeneration of electronic gas is given.
Earlier deduced correct formulas for
electric conductivity of quantum plasma is used.
For collisionless plasmas with the help
of kinetic approach the known formula of Landau diamagnetism is
dedu\-ced.

Thereby the answer to a question put in work \cite{Datta}
on dissipation influence on diamagnetism Landau is given.
Graphic research of properties of the magnetic
susceptibilities depending on dimensionless wave number,
chemical potential, frequency of oscillations of an electromagnetic field
and frequencies of collisions of particles of plasma is carried out.


\end{document}